\documentclass[11pt]{article}

\usepackage[margin=1in]{geometry}
\usepackage{amsmath,amssymb}
\usepackage{graphicx}
\usepackage{xcolor}
\usepackage{booktabs}
\usepackage{array}
\usepackage{enumitem}
\usepackage{microtype}
\usepackage{tikz}
\usetikzlibrary{arrows.meta,positioning,shapes.geometric,calc}
\usepackage[colorlinks=true,linkcolor=blue!60!black,citecolor=blue!60!black,urlcolor=blue!60!black]{hyperref}

\setlist{itemsep=2pt,topsep=4pt,parsep=0pt}

\newcommand{\failplausible}{\emph{fail-plausible}}
\newcommand{\classref}[1]{Class~#1}
\newcommand{\sysname}{\texttt{openclaw-model-bridge}}

\tikzset{
  box/.style={draw, rounded corners=2pt, align=center, font=\small,
              inner sep=4pt, minimum height=18pt},
  stage/.style={box, fill=gray!8},
  bad/.style={box, fill=red!8, draw=red!60!black},
  amp/.style={box, fill=orange!12, draw=orange!70!black},
  con/.style={box, fill=yellow!15, draw=yellow!50!black},
  good/.style={box, fill=green!8, draw=green!50!black},
  flow/.style={-{Stealth[length=2.2mm]}, thick},
  lbl/.style={font=\scriptsize\itshape, align=left},
}

\title{\textbf{When Errors Become Narratives: A Longitudinal Taxonomy of
Silent Failures in a Production LLM Agent Runtime}}

\author{%
  Wei Wu\thanks{Contact: wuweinanonuaa@gmail.com. \quad
  \textbf{Contributions \& AI disclosure:} the author designed, operates,
  and owns the system under study, provided domain expertise and all
  research direction, and made all editorial decisions; Claude
  (Anthropic), used through a coding-agent interface, served as AI
  engineering collaborator --- co-implementing the system and its defense
  framework, co-writing the incident postmortems under the author's
  protocol, preparing the quantitative inventory, and drafting this
  manuscript under the author's supervision. The arrangement is itself
  examined as a finding in \S\ref{sec:discussion}. All numbers are
  mechanically traceable to the public repository.}\\
  \small Independent researcher}

\date{June 2026 ~$\cdot$~ Draft v0.3 (LaTeX)}

\begin{document}
\maketitle

\begin{abstract}
\noindent
Large language model (LLM) agent systems are increasingly deployed as
long-running, autonomous runtimes---orchestrating scheduled jobs, calling
tools, maintaining memory, and pushing results to humans over messaging
channels. We present a longitudinal empirical study of \emph{silent
failures} in one such system: a personal-assistant agent runtime in
continuous production since March 2026, comprising roughly 40 scheduled
jobs, 8 LLM providers, a tool-governance proxy, and a knowledge-base
memory plane, defended by 4{,}286 unit tests and 827 declarative
governance checks. Over an eight-week window we documented 22 incidents
with full root-cause postmortems, within which a single
meta-pattern---\emph{a failure whose error signal never reaches a human
in actionable form}---manifested at least 28 times.

From these postmortems we derive a five-class, mechanism-oriented
taxonomy of silent failure: (A)~environment and platform quirks,
(B)~design-assumption mismatches, (C)~error swallowing and dilution,
(D)~chained hallucination and fabrication, and (E)~operational omission
and forensic blind spots. \classref{D} is, to our knowledge, specific to
LLM-based systems and the most dangerous: the system does not merely fail
to report an error---the LLM actively \emph{transforms} the error into
fluent, plausible narrative content delivered to the user. We term this
behavior \failplausible{}, and position it as the LLM-era escalation of
\emph{gray failure}'s differential observability: the observer is not
just blind, it is being convincingly lied to by the failure itself.

Three cross-cutting findings challenge common assumptions about agent
reliability engineering. First, discovery channel: roughly 70\% of silent
failures were ultimately caught by \emph{human user-view observation} of
system output, not by unit tests, health checks, or governance
audits---all of which stayed green through most incidents. Second, a
retrospective audit of 15 incidents found our declarative governance
layer had a \textbf{0\% ex-ante prevention rate but an 87\% ex-post
regression-blocking rate}---audits are regression engines, not prediction
engines. Third, incident latency (13 hours to 60 days of silence)
correlates with failure mechanism, not code complexity: the longest-lived
failures lived in the \emph{seams} between components---deployment
topology, cross-script contracts, observer--observed coupling---where, by
construction, no test runs. We describe the defense framework that
emerged (meta-rules, mechanized scanners, sabotage-validated invariants,
a declared-state convergence engine, and layered anti-fabrication
guards), and distill design principles for engineering LLM agent systems
whose failures are loud, attributable, and boring. All 22 postmortems,
the governance engine, and the defense framework are publicly available.
\end{abstract}

\section{Introduction}\label{sec:intro}

The reliability literature has long known that the most damaging failures
in production systems are not crashes. \emph{Gray
failures}~\cite{huang2017gray} degrade cloud systems while their failure
detectors report health; \emph{fail-slow} hardware~\cite{gunawi2018failslow}
throttles performance for hours before anyone suspects the disk. The
defining property of both is \textbf{differential observability}: the
application suffers, but the observer designed to notice does not.

LLM agent systems inherit this entire problem class---and then add a new
one. An agent runtime is a generator of fluent language. When an upstream
error leaks into its context window, the system's failure mode is not
silence; it is \emph{plausible speech}. In one incident we document
(\S\ref{sec:classd}), an HTTP~400 error page was captured into a cache by
a logging bug, and the downstream LLM---seeing error strings where
signals should be---confidently fabricated an industry analysis titled
around a ``Hugging Face platform crisis'' and pushed it to the user as a
routine insight digest. No detector fired. Every test was green. The
error had not disappeared; it had been \emph{narrated}.

We call this behavior \failplausible{}: a failure mode in which the
system transforms an internal error into coherent, contextually
appropriate, and false output. Fail-plausible is to LLM agent systems
what gray failure was to cloud infrastructure---the dominant,
hardest-to-see failure class---but it is strictly worse for the observer:
gray failure starves the failure detector of signal, while fail-plausible
feeds the human a counterfeit one.

This paper reports what eight weeks of documented production incidents in
a real, continuously operating LLM agent runtime taught us about silent
failures, in five contributions:

\begin{enumerate}

\item \textbf{A mechanism-oriented taxonomy} (\S\ref{sec:taxonomy}) of
five silent-failure classes derived from 22 fully documented production
incidents, each with a complete causal-chain postmortem. We classify by
\emph{failure mechanism} rather than by failure location, because the
same mechanism recurs across unrelated components, and a defense against
a mechanism immunizes a class rather than a file.

\item \textbf{The fail-plausible failure class} (\S\ref{sec:classd}),
with four documented incidents in which LLMs converted polluted context
(error logs, stale alerts, injected summaries) into confident
fabrications delivered to the user---including fabricated software
releases, fabricated platform crises, and fabricated remediation
instructions for the host operating system.

\item \textbf{Quantified cross-cutting findings} (\S\ref{sec:findings}):
incident latency distribution (13~hours to 60~days), discovery-channel
distribution ($\sim$70\% human user-view; unit tests close to 0\% for
this failure class), a three-layer root-cause structure
(trigger~/~amplifier~/~concealer) present in nearly every incident, and
the empirical observation that one meta-pattern manifested ${\geq}28$
times across all five classes---silent failure is a \emph{bug class}, not
a bug.

\item \textbf{An audited defense framework} (\S\ref{sec:defense}) and its
honest scorecard: a retrospective Q1/Q2/Q3 audit of 15 incidents showing
0\% ex-ante prevention but 87\% ex-post regression blocking, leading to
the position that \emph{audit is a regression engine, not a prediction
engine}; a three-step defense maturation path (point fix $\rightarrow$
meta-rule $\rightarrow$ mechanized scanner) with evidence that lessons
stopping at step one recur within days; and sabotage validation
(deliberately breaking the system to prove each guard actually fires).

\item \textbf{A complexity argument} (\S\ref{sec:discussion}): the
longest-latency failures did not live in complex components but in
\emph{seams}---between repository and deployment, between dev and target
OS, between declared state and runtime state, between observer and
observed. We argue that agent reliability engineering should optimize for
\emph{seam reduction} (representation unification, declared-state
convergence, read-only observers) over defense accretion, and report our
adoption of an explicit ``Sunset Law''---a standing rule that retiring
complexity outranks adding protection.

\end{enumerate}

Our study is a single-system longitudinal case study---the methodological
complement of recent horizontal studies that sample failures across many
frameworks from benchmark execution traces~\cite{cemri2025mast} or across
many customers from a provider's incident
database~\cite{ranganathan2025inference}. Horizontal studies establish
breadth; what they cannot see is the longitudinal texture of production
silent failure---multi-week latency, discovery channels, defense
evolution, recurrence of ``fixed'' lessons---because those only exist in
a system that runs continuously, pushes output to a real user, and keeps
complete postmortems. That texture is the subject of this paper.

\section{Background and Related Work}\label{sec:related}

\subsection{Silent failures in distributed systems}

Huang et~al.\ introduced \emph{gray failure} as the failure mode behind
most cloud incidents: components degrade while failure detectors report
health, formalized as \textbf{differential observability} between the
application's view and the observer's view~\cite{huang2017gray}. Gunawi
et~al.'s \emph{fail-slow at scale} study collected 101 (later 114)
incident reports of hardware performance faults across 12--14
institutions, documenting cascading root causes, fault conversion (one
form turning into another), and multi-hundred-hour diagnosis
costs~\cite{gunawi2018failslow}. Our study is methodologically downstream
of this tradition---taxonomy from production incident reports---but at a
different layer (an LLM agent runtime rather than cloud infrastructure or
hardware) and with a different lens (every incident in our corpus is a
\emph{silent} failure by selection; crashes that announced themselves
were handled as routine operations and did not generate postmortems).

Two of our five classes (C: error swallowing/dilution, E: operational
omission) will look familiar to readers of that literature.
\classref{D} will not.

\subsection{LLM agent failure studies}\label{sec:related-agents}

Cemri et~al.'s MAST is the first empirically grounded taxonomy of
multi-agent LLM system failures: 14 failure modes in 3 categories
(specification issues, inter-agent misalignment, task verification),
built from 1{,}600+ annotated execution traces across 7 frameworks with
strong inter-annotator agreement (Cohen's
$\kappa = 0.88$)~\cite{cemri2025mast}. MAST's unit of analysis is the
\emph{benchmark task trace}; failures are mostly visible in the trace as
task non-completion or wrong answers. Our unit of analysis is the
\emph{production incident}: failures that by definition did \textbf{not}
visibly fail any task, kept all detectors green, and were discovered
hours-to-months later. The two taxonomies are complementary: MAST
describes why agent collectives fail at tasks; we describe why an agent
\emph{system in operation} fails its operator without anyone noticing.

A recent provider-side study analyzed 156 high-severity LLM inference
incidents and derived a four-way operational taxonomy (infrastructure,
model configuration, inference engine, operational
failures)~\cite{ranganathan2025inference}. That work shares our
production-incident grounding but covers the \emph{inference service}
layer below agents, where failures manifest as availability and latency
violations---loud by nature. Ezell et~al.\ argue for structured incident
analysis of AI agents and propose what information agent incident reports
should contain~\cite{ezell2025incident}; our corpus can be read as an
existence proof of their program---22 reports collected under a fixed
postmortem protocol---and our experience adds a requirement their
framework should absorb: the report must record \emph{how long the
incident was silent and who finally noticed}, because for this failure
class those two fields carry most of the engineering signal
(\S\ref{sec:latency}--\S\ref{sec:discovery}). Related systems work
addresses exception handling in agentic workflows
(SHIELDA~\cite{zhou2025shielda}), incident response for agent safety
(AIR~\cite{xiao2026air}), and fault taxonomies for the Model Context
Protocol ecosystem~\cite{taraghi2026mcp,owotogbe2026mcpruntime}. None of
these focuses on the silent/fail-plausible class, the longitudinal
latency-and-discovery question, or defense-framework evolution under real
recurrence pressure.

\subsection{Hallucination research}

Hallucination is usually studied as a \emph{model} property---ungrounded
generation measured against references, with taxonomies of
intrinsic/extrinsic and factuality/faithfulness variants, detection
benchmarks, and model-side mitigations~\cite{huang2023hallucination}. Our
\classref{D} reframes it as a \emph{systems} property: in 4/4 documented
fabrication incidents the model behaved exactly as trained (fluent
completion over its context); the failure was that \textbf{the system
delivered polluted context} (error logs captured by command substitution
into a cache; stale alert messages persisted into the chat history;
injected cross-day summaries without provenance marking). The defense
consequently is not model-side but system-side: context hygiene (stderr
discipline, alert stripping before context assembly), provenance labeling
(source-credibility tiers), and layered anti-fabrication guards with
explicit, literal prohibitions. This systems view of
hallucination---\emph{garbage context in, confident narrative out}---is a
distinct contribution relative to model-centric hallucination work.

\subsection{Operational practice: SRE, chaos engineering, and incident
studies}

Site Reliability Engineering codified error budgets, postmortem culture,
and the principle that operational knowledge must be engineered rather
than remembered~\cite{beyer2016sre}; our convergence engine
(\S\ref{sec:defense}) is that principle applied to an agent runtime's
declared state. Chaos engineering established deliberate fault injection
as the way to gain confidence in a system's
resilience~\cite{basiri2016chaos}; our framework applies the same
epistemology one level up---\emph{sabotage validation}
(\S\ref{sec:defense}) injects violations to gain confidence in the
\textbf{guards}, because in a silent-failure regime an unvalidated
detector is indistinguishable from a vacuous one, a lesson we learned
from 67 checks that had silently executed empty strings for months.
Large-scale incident studies such as Ghosh et~al.'s analysis of 152
severe incidents in a production cloud service~\cite{ghosh2022fight}
established the empirical template---incident corpus, root-cause and
mitigation coding, automation gap analysis---that we instantiate for an
agent runtime; the variable our setting adds is that the system under
study \emph{speaks}, which changes both what failure looks like
(\S\ref{sec:classd}) and what detection requires
(\S\ref{sec:discovery}).

\section{System Context and Methodology}\label{sec:method}

\subsection{The system under study}\label{sec:system}

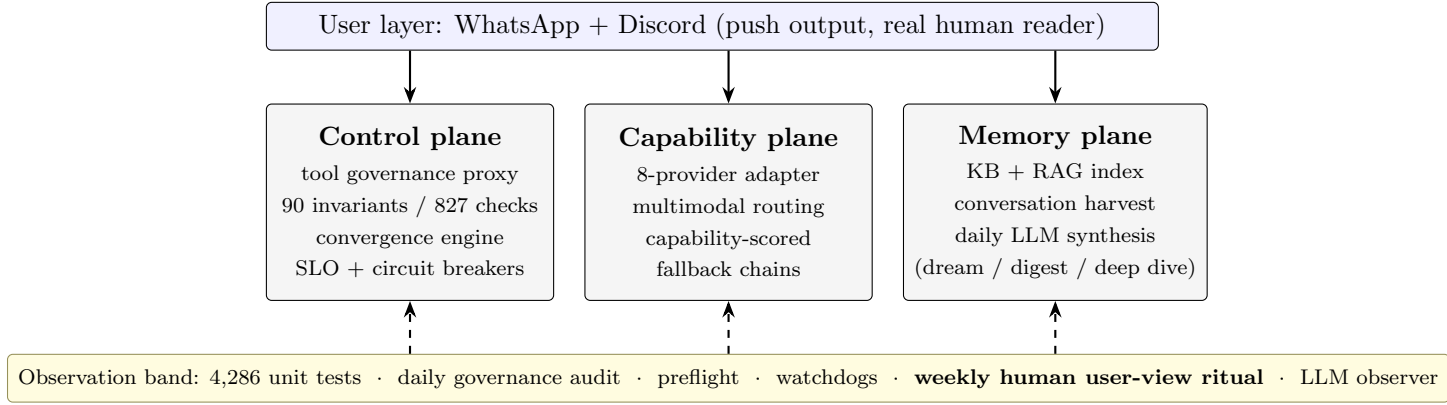
\begin{figure}[t]
\centering
\begin{tikzpicture}[node distance=4mm and 5mm]
  \node[box, fill=blue!6, minimum width=118mm] (user)
    {User layer: WhatsApp + Discord (push output, real human reader)};
  \node[stage, minimum width=38mm, minimum height=26mm, below=7mm of user.south west, anchor=north west, align=center]
    (ctrl) {\textbf{Control plane}\\[1pt]\scriptsize tool governance proxy\\\scriptsize 90 invariants / 827 checks\\\scriptsize convergence engine\\\scriptsize SLO + circuit breakers};
  \node[stage, minimum width=38mm, minimum height=26mm, right=4mm of ctrl]
    (cap) {\textbf{Capability plane}\\[1pt]\scriptsize 8-provider adapter\\\scriptsize multimodal routing\\\scriptsize capability-scored\\\scriptsize fallback chains};
  \node[stage, minimum width=38mm, minimum height=26mm, right=4mm of cap]
    (mem) {\textbf{Memory plane}\\[1pt]\scriptsize KB + RAG index\\\scriptsize conversation harvest\\\scriptsize daily LLM synthesis\\\scriptsize (dream / digest / deep dive)};
  \node[con, minimum width=118mm, below=7mm of cap.south, anchor=north]
    (obs) {\scriptsize Observation band: 4{,}286 unit tests ~$\cdot$~ daily governance audit ~$\cdot$~ preflight ~$\cdot$~ watchdogs ~$\cdot$~
           \textbf{weekly human user-view ritual} ~$\cdot$~ LLM observer};
  \draw[flow] (user.south -| ctrl.north) -- (ctrl.north);
  \draw[flow] (user.south -| cap.north) -- (cap.north);
  \draw[flow] (user.south -| mem.north) -- (mem.north);
  \draw[flow, dashed] (obs.north -| ctrl.south) -- (ctrl.south);
  \draw[flow, dashed] (obs.north -| cap.south) -- (cap.south);
  \draw[flow, dashed] (obs.north -| mem.south) -- (mem.south);
\end{tikzpicture}
\caption{The system under study: a three-plane agent runtime with an
observation band beneath it. The incidents in this paper are, by
selection, the ones that crossed all planes \emph{and} the observation
band without raising an actionable signal---most were finally caught at
the very top, by the human reading pushed output
(\S\ref{sec:discovery}).}
\label{fig:planes}
\end{figure}

The subject system (public repository \sysname{}) is a two-layer
middleware that connects self-hosted and commercial LLMs to OpenClaw, an
open-source personal-agent framework bridging WhatsApp and Discord. It
has been in continuous production on a single macOS host since early
March 2026. Architecturally it follows a three-plane design
(Fig.~\ref{fig:planes}):

\begin{itemize}
\item \textbf{Control plane} --- a tool-governance proxy (tool filtering
to a hard cap, schema repair, alert-context stripping, custom tool
interception), a declarative governance engine (90~invariants,
827~checks, 23~meta-rules, 14~mechanized discovery scanners, daily audit
cron), SLO monitoring, circuit breakers, and a convergence engine that
machine-synchronizes declared state (job registry, provider registry,
service registry) to runtime state (crontab, fallback chains, launchd).
\item \textbf{Capability plane} --- an adapter routing chat/completions
across 8 providers (self-hosted Qwen3-235B primary, Qwen2.5-VL multimodal
route, plus commercial fallbacks) with capability-scored automatic
fallback chains.
\item \textbf{Memory plane} --- a knowledge base ($\sim$1{,}100+ notes,
local-embedding RAG index), multimodal media index, conversation-harvest
pipeline, and daily LLM-driven synthesis jobs (``dream'', ``evening
digest'', ``deep dive'') that push insights to the user.
\end{itemize}

Scale snapshot at the study cutoff (2026-06-11): ${\sim}40$ registered
scheduled jobs across two cron schedulers; 8 LLM providers; 3 supervised
long-running services; 4{,}286 unit tests in 121 suites; 22 published
incident postmortems. One human operator (the system owner) and one AI
engineering collaborator (Claude, used through a coding-agent interface)
develop and operate the system; the agent runtime itself is powered by
separate models (Qwen3-class), making this---to our knowledge---also a
data point on AI-assisted operation of an AI system.

\subsection{Incident corpus and postmortem protocol}\label{sec:protocol}

The corpus is every incident between 2026-04-09 and 2026-06-02 that
(a)~reached production, (b)~had a silent phase---a period in which the
system was failing or had failed while all automated indicators stayed
green---and (c)~was closed with a full postmortem. 22 incidents qualify.
Postmortems follow a mandatory in-repo protocol (the
``exception-analysis constitution'') requiring, before any fix:

\begin{enumerate}
\item \textbf{Full causal-chain diagram} --- timeline $\times$ layer
$\times$ logic $\times$ architecture, from upstream trigger to
user-perceived symptom, with code-branch truth values annotated;
\item \textbf{Three-layer root cause} --- \emph{trigger} (what external
event ignited it), \emph{amplifier} (what architectural flaw spread it),
\emph{concealer} (what absence hid it until a human noticed);
\item \textbf{Timeline reconstruction} with per-minute precision where
logs allow;
\item \textbf{Condition-combination analysis} --- why now and not before:
which set of individually benign, often long-latent conditions first
co-occurred;
\item \textbf{Feeding the governance ontology} --- new invariants,
meta-rule candidacy, catalog entry.
\end{enumerate}

The protocol was itself a product of early incidents (it was adopted
after the second incident in the corpus) and was applied retroactively to
the first two. Each postmortem is a standalone document in the public
repository, and the consolidated catalog is the canonical index from
which this paper's taxonomy is built.

\subsection{Taxonomy construction}\label{sec:construction}

We classify by \textbf{mechanism} (how the error evaded observation)
rather than by \textbf{location} (which job or file failed). The
location-oriented draft of the catalog was produced first and discarded:
the same mechanism (e.g., positional parsing of LLM output; copy-pasted
error-suppression idioms) recurred across unrelated jobs, so location
classes had no predictive or defensive value, while one mechanism-level
defense (e.g., a repo-wide scanner for the suppression idiom) immunized
every location at once. Classes were derived iteratively: each new
postmortem was matched against existing classes' ``true root cause''
column; two consecutive non-matches forced a class split or a new class.
The five-class scheme has been stable since mid-May 2026 over the last 9
incidents.

Two limitations are flagged here and expanded in
\S\ref{sec:threats}: classification was performed by the two system
operators (human + AI collaborator) without independent annotators, so we
report no inter-annotator agreement; and selection requires a
\emph{completed} postmortem, so failures still silent today are by
construction absent (we discuss this surviving-silence bias in
\S\ref{sec:threats}).

\subsection{The meta-pattern counter}\label{sec:counter}

Early in the study we adopted a meta-rule, \textbf{MR-4
``silent-failure-is-a-bug''}, and began counting its
\emph{manifestations}---distinct incidents or sub-events in which an
error signal existed somewhere in the system but never reached a human in
actionable form. The counter reached ${\geq}28$ across the 22 incidents
(several incidents contain multiple distinct manifestations; numbering
has deliberate gaps where manifestations were recorded inside hotfix
notes). We report the counter not for its precision but for its shape:
manifestations continued to appear at a roughly constant rate even as
defenses accumulated---in \emph{new forms} each time
(\S\ref{sec:bugclass}). This is the empirical basis for treating silent
failure as a bug class rather than a finite list of bugs.

\section{A Taxonomy of Silent Failures}\label{sec:taxonomy}

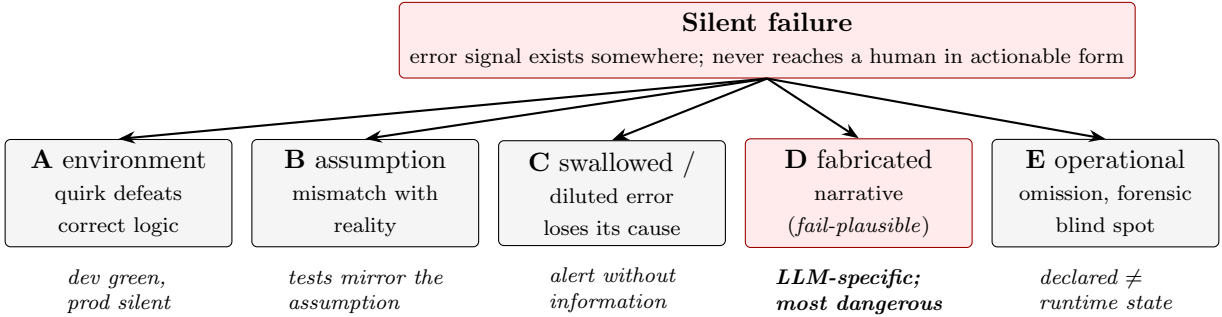
\begin{figure}[t]
\centering
\begin{tikzpicture}[node distance=3mm and 2.5mm]
  \node[bad, minimum width=72mm] (root)
    {\textbf{Silent failure}\\\scriptsize error signal exists somewhere; never reaches a human in actionable form};
  \node[stage, below left=8mm and 22mm of root, minimum width=30mm] (a)
    {\textbf{A} environment\\\scriptsize quirk defeats\\\scriptsize correct logic};
  \node[stage, right=2.5mm of a, minimum width=30mm] (b)
    {\textbf{B} assumption\\\scriptsize mismatch with\\\scriptsize reality};
  \node[stage, right=2.5mm of b, minimum width=30mm] (c)
    {\textbf{C} swallowed /\\\scriptsize diluted error\\\scriptsize loses its cause};
  \node[bad, right=2.5mm of c, minimum width=30mm] (d)
    {\textbf{D} fabricated\\\scriptsize narrative\\\scriptsize (\failplausible{})};
  \node[stage, right=2.5mm of d, minimum width=30mm] (e)
    {\textbf{E} operational\\\scriptsize omission, forensic\\\scriptsize blind spot};
  \foreach \x in {a,b,c,d,e} \draw[flow] (root.south) -- (\x.north);
  \node[lbl, below=1.5mm of a] {dev green,\\prod silent};
  \node[lbl, below=1.5mm of b] {tests mirror the\\assumption};
  \node[lbl, below=1.5mm of c] {alert without\\information};
  \node[lbl, below=1.5mm of d] {\textbf{LLM-specific;}\\\textbf{most dangerous}};
  \node[lbl, below=1.5mm of e] {declared $\neq$\\runtime state};
\end{tikzpicture}
\caption{Mechanism-oriented taxonomy. Classes C and E have direct
ancestry in the gray-failure / fail-slow literature; \classref{D} is
specific to systems that generate language.}
\label{fig:tree}
\end{figure}

Table~\ref{tab:classes} and Fig.~\ref{fig:tree} summarize the five
classes. Full per-incident detail is in the public catalog; here we
define each class, give its mechanism signature, and narrate one or two
representative incidents.

\begin{table}[t]
\centering\small
\caption{Five classes over 22 incidents (+ sub-events).}
\label{tab:classes}
\begin{tabular}{@{}p{31mm}p{40mm}cp{17mm}p{36mm}@{}}
\toprule
\textbf{Class} & \textbf{Mechanism} & \textbf{Inc.} &
\textbf{Silence} & \textbf{Defining property} \\
\midrule
A --- Environment / platform quirk & Logic correct; runtime
environment's implicit behavior defeats it & 1 (+6) & hours--weeks &
dev always green; only target OS/client reveals \\
\addlinespace[2pt]
B --- Design-assumption mismatch & Code assumes a deployment topology /
contract / input shape that reality violates & 4 & days & unit tests
cover the assumption, not reality \\
\addlinespace[2pt]
C --- Error swallowing \& dilution & Error occurs, is eaten by a layer
or stripped of cause across layers & 5 & hours--days & the alert that
arrives carries no usable information \\
\addlinespace[2pt]
D --- Chained hallucination \& fabrication & LLM converts polluted
context into confident false output & 4 & hours--days &
\failplausible{}: user receives counterfeit health \\
\addlinespace[2pt]
E --- Operational omission \& forensic blind spot & Deployment /
registration step skipped; or the forensic tool itself is blocked and
reads as ``normal'' & 8 & days--\textbf{60 days} & declared state
$\neq$ runtime state; diagnosis instruments lie \\
\bottomrule
\end{tabular}
\end{table}

\subsection{Class A --- Environment and platform quirks}\label{sec:classa}

\emph{The logic is right; the environment's implicit behavior is not what
anyone assumed.} The development environment (Linux container, root, GNU
userland, bash~5) is systematically more permissive than the production
target (macOS, bash~3.2 as \texttt{/bin/bash}, BSD userland, zsh
interactive shell, sandboxed cron). Documented sub-events include:
bash~3.2 not propagating ERR traps into functions without
\texttt{set~-E}, silently disarming a watchdog's self-alarm; BSD awk
aborting on invalid UTF-8 multibyte sequences, which---combined with
\texttt{set~-e} and \texttt{pipefail}---killed the \emph{monitoring}
script itself for 7~days (\S\ref{sec:classe} overlaps); the absence of
GNU \texttt{timeout} on stock macOS turning a defensive wrapper into a
universal ``tool unavailable'' failure; CJK full-width punctuation
adjacent to unbraced shell variables parsing as part of the variable
name; and a messaging client folding long messages at an undocumented
${\sim}4{,}000$-character threshold, silently changing the user-visible
shape of every long push.

The class signature is \textbf{green dev, silent prod}, and its defense
is necessarily mechanized: a cross-OS quirk scanner now encodes six known
quirk patterns as repository-wide checks, and a meta-rule requires
framework-level fixes to be validated \emph{on the target environment},
after a fix validated only in dev shipped broken twice in one day.

\subsection{Class B --- Design-assumption mismatches}\label{sec:classb}

\emph{The code is internally consistent with an assumption; production
violates the assumption.} Representative incident: a metadata-resolution
function shipped with three path candidates for locating a registry
file, all of which missed the file's actual production
location---because the deployment pipeline placed it under a path none of
the candidates covered. The component fell back, silently, to an
unfiltered mode for \textbf{five days}; its own log line announcing the
fallback scrolled past unread, and a companion feature (writing health
scores into shared state) silently never executed at all. Unit tests
were green throughout: they tested the resolution \emph{logic}, with
fixtures laid out according to the same wrong assumption.

A second representative: an LLM-output parser indexed lines positionally
(\texttt{lines[i+1]}, \texttt{lines[i+2]}, stride~3). The model
occasionally omits one line---instruction-following is a distribution,
not a contract---after which every subsequent field shifted one slot,
and users received messages whose ``title'' field contained a literal
separator string. The mechanism generalizes: \textbf{any positional
parse of LLM output is a latent \classref{B} failure}, which we later
froze into a meta-rule (parsers must be key-based, never positional) and
a scanner.

The class signature is \emph{tests mirror the assumption rather than the
caller}: in a later incident the test suite constructed inputs with a
heading format the production caller never produces, passing 12 tests
while production emitted 3 message windows instead of 4. Defense:
explicit cross-script contracts, deployment-layout testing on target,
scanners that walk every path-resolution function and assert the
production-canonical candidate is present.

\subsection{Class C --- Error swallowing and dilution}\label{sec:classc}

\emph{The error happens and is reported---into a void.} Three mechanism
variants:

\paragraph{Swallowing.} A summary function counted only
\texttt{status=="fail"} results; invariants whose check raised an
\emph{exception} (\texttt{status=="error"}) vanished, and the governance
audit printed ``all invariants hold'' over a pile of dead checks. The
observer had no observer---we return to this as the \emph{observer's own
blind spot} in \S\ref{sec:defense}. In a later echo of the same class, a
governance executor read check code from one YAML field while 67 checks
(across 21 invariants) had their code in a differently named field: all
67 executed \texttt{exec("")}---vacuous passes---for months, and were
discovered only when a \emph{new} invariant's sabotage validation refused
to fail.

\paragraph{Dilution.} An evening digest failed with ``HTTP 502: Bad
Gateway'' for two consecutive days. The true cause---the fallback
provider's free-tier quota exhausted by daytime jobs, after the primary's
circuit breaker opened---was present in the upstream response body, which
the adapter wrapped into a 502, whose body the proxy never read
(\texttt{str(e)} only), whose remnant the client reduced to
\texttt{f"HTTP \{code\}: \{reason\}"}. Three hops, each individually
reasonable, each stripping cause; the alert that reached the human
contained zero actionable bits. The meta-rule that emerged---\emph{error
chains must preserve upstream cause across layers}---is the LLM-pipeline
restatement of a classic distributed-tracing lesson, but with a twist: in
agent systems the ``client'' of the error is often another LLM prompt, so
a diluted error is one step from becoming \classref{D} fabrication input.

\paragraph{Amplified swallowing.} An automated batch tool injected an
environment-variable read into 8 job scripts without injecting the
corresponding \texttt{import}---a \texttt{NameError} in all 8, caught by
a fail-open guard, which dutifully skipped the new validation logic in
all 8 while every test and syntax check passed. Automation is a bug
amplifier: a single wrong design decision propagates to $N$ sites with
machine efficiency, and \texttt{bash~-n} cannot see inside a Python
heredoc.

\subsection{Class D --- Chained hallucination and fabrication
(fail-plausible)}\label{sec:classd}

\begin{figure}[t]
\centering
\begin{tikzpicture}[node distance=2.6mm]
  \node[stage, minimum width=96mm] (s1)
    {scraped content with isolated UTF-16 surrogate (U+D800--DFFF)};
  \node[bad, minimum width=96mm, below=of s1] (s2)
    {\texttt{json.dump} raises mid-write $\Rightarrow$ request body truncated\\
     \scriptsize adapter: \texttt{json.loads} fails $\Rightarrow$ HTTP 400 ``Bad JSON'' (456-byte HTML error page)};
  \node[amp, minimum width=96mm, below=of s2] (s3)
    {\textbf{Amplifier:} \texttt{log()} prints the error dump to \emph{stdout};\\
     \scriptsize \texttt{signals=\$(llm\_call \dots)} command substitution captures stdout $\Rightarrow$ error log becomes the ``signal'' payload};
  \node[con, minimum width=96mm, below=of s3] (s4)
    {\textbf{Concealer:} cache file := ``\dots Error code: 400 Bad JSON\dots'';\\
     \scriptsize non-empty check passes; status file reports ok; no detector inspects content semantics};
  \node[bad, minimum width=96mm, below=of s4] (s5)
    {reduce-step LLM reads cache as cross-domain ``signals''\\
     \scriptsize $\Rightarrow$ fluent synthesis: \emph{``Hugging Face platform crisis''} analysis};
  \node[box, fill=blue!6, minimum width=96mm, below=of s5] (s6)
    {pushed to the user via WhatsApp/Discord as a routine insight digest\\
     \scriptsize \textbf{no alarm} --- discovered by the user noticing a thematic incoherence};
  \node[good, minimum width=96mm, below=of s6] (s7)
    {fix that severed the chain: \texttt{log() \{ echo \dots\ >\&2; \}} \quad (one redirection)\\
     \scriptsize defense-in-depth added: surrogate sanitization $\cdot$ encoding error policy $\cdot$ anti-pollution prompt guards};
  \foreach \i/\j in {s1/s2,s2/s3,s3/s4,s4/s5,s5/s6} \draw[flow] (\i.south) -- (\j.north);
  \draw[flow, dashed, green!40!black] (s6.south) -- (s7.north);
\end{tikzpicture}
\caption{The D1 pollution chain: from one malformed byte to a fabricated
industry analysis. Every component behaves as designed. Trigger,
amplifier, and concealer layers (\S\ref{sec:tac}) are marked.}
\label{fig:chain}
\end{figure}
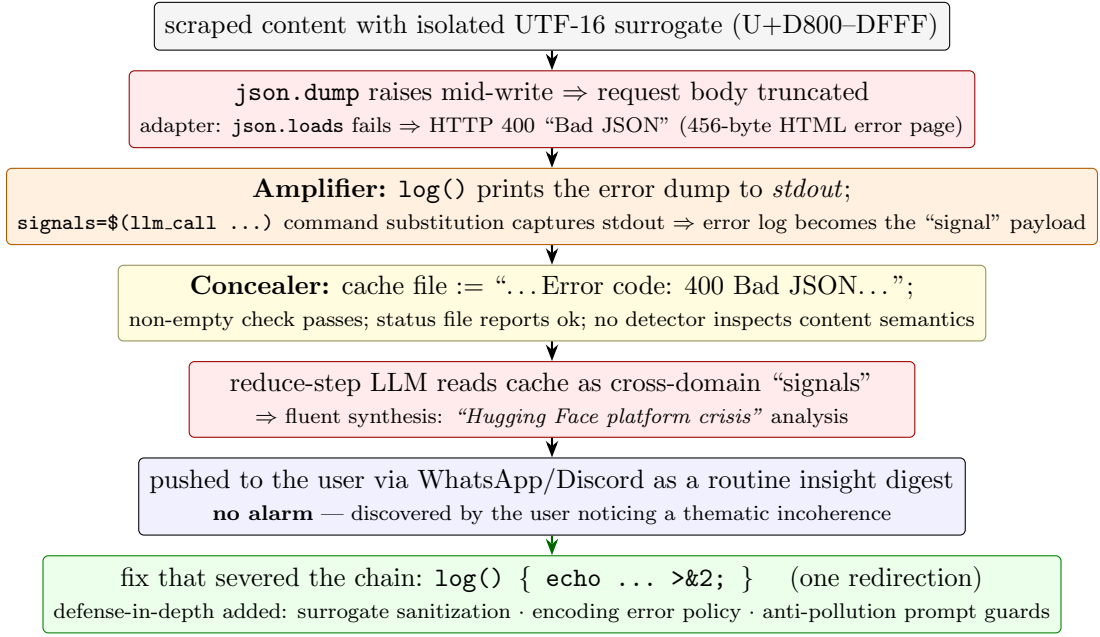

\emph{The most dangerous class, and the one without precedent in the
gray-failure literature.} The error is not suppressed; it is
\textbf{transformed}. Four incidents:

\paragraph{D1 --- The fabricated platform crisis.} A nightly
knowledge-synthesis job (``dream'') collects signals from ${\sim}290$
notes via map-reduce LLM calls. A Unicode surrogate in scraped content
made \texttt{json.dump} raise mid-write, producing a truncated request
body, a 400 from the adapter---and here the chain turns: the map step's
logging function wrote diagnostics to \textbf{stdout}, the caller
captured stdout by command substitution as the \emph{signal payload}, and
the cache filled with \texttt{HTTP Error 400: Bad JSON} strings. The
reduce-step LLM, prompted to find cross-domain signals, did exactly what
language models do with anomalous-but-thematic context: it composed a
confident analysis of a ``Hugging Face platform crisis''---platform
trouble being the most probable narrative shell for error-code
vocabulary---and the system pushed it to the user as a routine insight
digest. Every component succeeded; the pipeline laundered an encoding bug
into industry analysis (Fig.~\ref{fig:chain}). The structural fix was
four lines deep in defense (stderr discipline, surrogate sanitization,
encoding error policy, anti-pollution prompt guards), but the
load-bearing one was a single \texttt{>\&2}: \textbf{one redirection
operator severed the entire hallucination chain.}

\paragraph{D2 --- The fabricated remediation.} A system alert pushed by a
watchdog was persisted into the chat session history as an ordinary
assistant message. Thirty-six minutes later the user asked an unrelated
architecture question; the model, attending across the contaminated
context, replied that it had ``received the system alert follow-up task''
and instructed the user to grant Full Disk Access to a cron binary in
macOS System Preferences---instructions that were not only off-topic but
\emph{technically fabricated} for the scenario. Alert traffic and
conversation are different speech acts; letting them share a context
window invites the model to weave them into one narrative. (A later
forensic twist worth reporting: weeks afterward, an unrelated 60-day
investigation (\S\ref{sec:classe}) established that the \emph{general
remediation direction} the model fabricated---FDA for cron-derived
processes---was coincidentally relevant to a real problem elsewhere in
the system. Fabrication is not refuted by occasional accuracy; it is
defined by absence of grounding.)

\paragraph{D3 --- The fabricated success.} A weekly review job whose LLM
call failed fell back to a mechanical line-filter that emitted leftover
container headings as if they were review content, and wrote
\texttt{"llm": true} into its status file unconditionally. Six co-located
issues each made the others harder to see; the user-visible artifact
looked plausible enough to pass casual inspection for weeks. Fabrication
does not require a model: \textbf{a fallback path that manufactures
plausible-shaped output is a hallucination implemented in shell.}

\paragraph{D4 --- The fabricated release.} An evening digest, given a
list of the day's high-alignment papers as context enrichment, inferred
that the user's project ``must have shipped'' and announced a community
release of an internal version number that exists only in the project
changelog---inventing a source tag for it. The injection of \emph{true
but unlabeled} context produced false attribution: provenance-free
enrichment is fabrication fuel.

\medskip
The common structure is a \textbf{pollution chain}: a \classref{A}/B/C
failure deposits non-signal content where a downstream LLM expects
signal; the LLM performs its function---fluent, coherent completion---and
the output inherits the \emph{form} of health with the \emph{content} of
failure. Defense is therefore system-side context hygiene, applied at
every link: stderr discipline so diagnostics can never enter data
channels; alert stripping before context assembly;
provenance/credibility labeling of injected content; and a six-level
cumulative ladder of anti-fabrication prompt guards (from ``never invent
facts'' through literal prohibited phrases extracted from actual
incidents, e.g.\ the exact fabricated release string), shared as a single
imported module across all nine LLM-calling jobs rather than copy-pasted.
We make no claim that prompt guards are sufficient---they are the
\emph{last} layer, behind hygiene and provenance, and
\S\ref{sec:maturation} shows why layered placement matters more than any
single guard.

\subsection{Class E --- Operational omission and forensic blind
spots}\label{sec:classe}

\emph{The code is right; an operational step never happened---or the
diagnostic instrument itself is compromised.} This is the largest class
(8 incidents) and contains the longest silences.

\paragraph{Omission.} A new daily-analysis job was fully implemented,
tested, registered in the job registry, and deployed---and never ran,
because the final step (writing the crontab line) was a human memory
item. Three independent small bugs conspired to hide it: the preflight
checker grepped for only one of two registry-drift warning strings; the
crontab helper did not check its own install exit code and compared
counts with \texttt{<} instead of equality (reporting a green check on a
rejected install); and the job's absence produced no log to scan. The
incident generalized into the system's largest architectural correction:
\textbf{declared state must converge to runtime state via machine, not
memory}, implemented as a convergence engine that diffs registry
declarations against observed crontab/launchd/provider state on every
audit, with a staged escalation path (alert-only $\rightarrow$ dry-run
$\rightarrow$ machine-sync) and, after a one-week zero-drift observation
window per spec, automatic synchronization.

\paragraph{The reserved-file incident} deserves narration for its shape.
The agent, completing an alert-handling task, wrote ``task complete''
notes into a file named \texttt{HEARTBEAT.md} in its workspace---to the
agent, a scratch filename; to the runtime, a \emph{reserved control file}
whose non-empty content activates a heartbeat protocol that instructs the
model to reply with a bare acknowledgment token if nothing needs
attention. The gateway then strips that token from outbound messages.
Result: for 13~hours, every user message received a reply of
\texttt{HEARTBEAT\_OK}, stripped to nothing in transit---\textbf{total
silence, with every component functioning exactly as designed.} The model
had, in effect, been handed a pen that doubled as the system's mute
button. The general lesson became a meta-rule: any file path carrying
special runtime semantics must be unwritable by the LLM's generic file
tools (write attempts are intercepted at the proxy and rewritten to a
comments-only placeholder).

\paragraph{Forensic blind spots.} The longest silence in the corpus---60
days---was an external-SSD backup path failing with EPERM. Six successive
hypotheses (filesystem format, ownership UIDs, ACLs, daemon contention,
snapshot locks, physical disconnection) were each falsified by data over
multiple weeks; the breakthrough came only from the OS's own audit log
(\texttt{log show}), which revealed macOS TCC sandbox denials:
cron-derived processes lack Full Disk Access by default. The deeper
finding is methodological: during those weeks, the \emph{forensic
collectors themselves} (lsof, ACL listing, snapshot enumeration) were
being silently denied by the same sandbox and returning empty
output---which the diagnosis pipeline recorded as ``normal/empty''.
\textbf{An instrument that cannot distinguish ``nothing there'' from ``I
was not allowed to look'' will manufacture false reassurance}; all
collectors now capture stderr separately and tag
\texttt{[sandbox\_denied]} as a first-class observation. The companion
lesson---every falsified hypothesis was answered by \emph{adding a
forensic dimension}, never by speculative code change---is the discipline
we found most transferable.

Also in this class: a 9-hour total gateway outage whose three independent
alarms each failed (a quiet-hours filter suppressing both channels
including the one designed for emergencies; a keepalive that logged WARN
without alerting; a restart script that reported success without
verifying health)---yielding the meta-rule that \textbf{an alert path
must not depend on the failing subject} (gateway-down alerts must travel
a channel the gateway cannot take down); and the watchdog that died of a
\classref{A} quirk and stayed dead for 7~days because nothing watched the
watcher (now: ERR-trap self-alarms plus a heartbeat canary file that an
independent job can age-check).

\section{Cross-Cutting Findings}\label{sec:findings}

\subsection{Latency: how long silence lasts}\label{sec:latency}

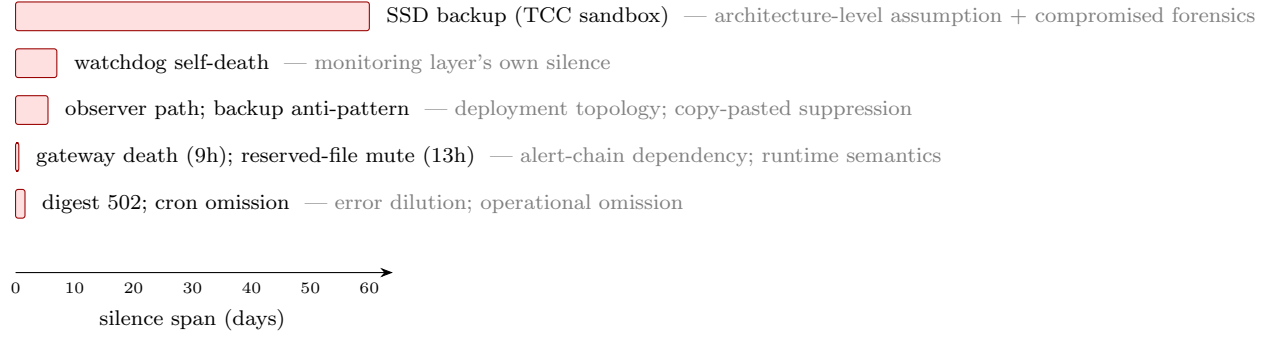
\begin{figure}[t]
\centering
\begin{tikzpicture}[xscale=0.078]
  \foreach \y/\len/\name/\mech in {
      0/60/SSD backup (TCC sandbox)/architecture-level assumption + compromised forensics,
      -1/7/watchdog self-death/monitoring layer's own silence,
      -2/5.5/observer path; backup anti-pattern/deployment topology; copy-pasted suppression,
      -3/0.55/gateway death (9h); reserved-file mute (13h)/alert-chain dependency; runtime semantics,
      -4/1.6/digest 502; cron omission/error dilution; operational omission}{
    \draw[fill=red!12, draw=red!60!black, rounded corners=1pt]
      (0,\y*0.62) rectangle (\len,\y*0.62+0.38);
    \node[anchor=west, font=\scriptsize] at (\len+1.2,\y*0.62+0.19)
      {\name\ \ \textcolor{gray}{---\ \mech}};
  }
  \draw[-{Stealth}, thin] (0,-3.2) -- (64,-3.2);
  \foreach \x in {0,10,...,60} \node[font=\tiny, below] at (\x,-3.2) {\x};
  \node[font=\scriptsize, below=3.5mm] at (30,-3.2) {silence span (days)};
\end{tikzpicture}
\caption{Silence span by incident group. Latency tracks the mechanism
layer, not code complexity: the long tail lives where no test runs.}
\label{fig:latency}
\end{figure}

Latency tracks the \textbf{mechanism layer, not code complexity}
(Fig.~\ref{fig:latency}). Code-level bugs die young: unit tests and the
next run catch them. What survives for weeks lives where no test
runs---in deployment topology, OS policy, monitoring-of-monitoring, and
the gap between declared and runtime state. Latency is therefore a
\emph{measure of observational distance}: the further a mechanism sits
from any existing observer, the longer it lives. We suggest
silence-latency percentiles as a reportable reliability metric for agent
systems, complementary to MTTR---our experience is that for silent
failures, time-to-\emph{detect} dominates time-to-repair by one to two
orders of magnitude.

\subsection{Discovery: who finally notices}\label{sec:discovery}

\begin{table}[t]
\centering\small
\caption{Discovery channels across the corpus (qualitative shares).}
\label{tab:discovery}
\begin{tabular}{@{}p{52mm}p{18mm}p{52mm}@{}}
\toprule
\textbf{Channel} & \textbf{Share} & \textbf{Notes} \\
\midrule
\textbf{Human user-view} (reading actual pushed output; weekly
observation ritual) & \textbf{$\sim$70\%} & ``this digest looks
shallow'', ``why two windows?'', ``didn't receive yesterday's
analysis'' \\
\addlinespace[2pt]
Target-environment execution (dev green, prod reveals) & high & all
\classref{A}; several \classref{B} \\
\addlinespace[2pt]
Self-observation (governance auditing governance; observer critiquing
output) & rising & added mid-study; caught its own executor bug \\
\addlinespace[2pt]
Unit tests / preflight & $\approx$0 for this corpus & by selection:
anything they caught never became a silent incident \\
\bottomrule
\end{tabular}
\end{table}

The last row of Table~\ref{tab:discovery} is partly tautological---our
corpus selects for what tests missed---but the magnitude of the first row
is not. The system runs 4{,}286 tests, 827 governance checks, and a
19-point preflight, all green through most of these incidents; the single
most productive detector of silent failure was \textbf{a human looking at
the product as a user}. We institutionalized this as a weekly 30-minute
observation ritual (no coding allowed; four dimensions: alert noise, push
latency, information density, response quality)---and it continued to
out-detect the automated stack. For practitioners, the implication is
blunt: \emph{user-view observation is a first-class observability signal
and deserves calendar time}; for researchers, the open problem is
mechanizing even part of what the human eye does here. The LLM-as-judge
literature gives grounds for optimism---strong LLM judges reach
${>}80\%$ agreement with human preference on open-ended response
quality~\cite{zheng2023judge}---and our own step in that direction, a
daily LLM ``observer'' that critiques yesterday's outputs against quality
heuristics, found real regressions (including a fabrication, and
including two bugs in \emph{itself}). The qualifier our experience adds
to that literature: an LLM judging a \emph{system's output for silent
failure} is itself an LLM component of that system, inheriting every
class in this taxonomy---ours shipped with a \classref{B} path bug and a
sampling artifact that made it hallucinate a truncation---so the judge
needs the same governance, provenance hygiene, and sabotage validation as
the components it judges.

\subsection{The trigger--amplifier--concealer structure}\label{sec:tac}

Nearly every postmortem decomposed into three causally distinct layers:

\begin{center}\small
\begin{tabular}{@{}lll@{}}
\toprule
\textbf{Layer} & \textbf{Role} & \textbf{Examples} \\
\midrule
trigger & the external spark & a surrogate byte; an omitted LLM output
line; a transient EPERM \\
amplifier & the architectural flaw that spreads it & stdout logging into
command substitution; \\
& & positional parsing; 18 copy-pasted suppression idioms \\
concealer & the absence that hides it & status file lying ``ok'';
fail-open guard; quiet-hours filter; \\
& & forensic tool silently denied \\
\bottomrule
\end{tabular}
\end{center}

The practical force of the decomposition is prescriptive: \textbf{a fix
that addresses only the trigger is cosmetic}. Triggers are unbounded (the
environment will always produce another malformed byte, another quirk);
amplifiers and concealers are finite and owned by the architecture. Every
recurrence in our corpus (\S\ref{sec:bugclass}) traces to a fix that
stopped at the trigger. Conversely, the highest-leverage single fixes
were amplifier-level (one \texttt{>\&2}; one shared helper replacing 20
idiom copies) and concealer-level (fail-loud with cause; forensic stderr
tagging).

\subsection{Silent failure is a bug class, not a bug}\label{sec:bugclass}

The MR-4 manifestation counter (\S\ref{sec:counter}) kept advancing at a
roughly constant rate across the study---but never twice in the same
form. Early manifestations were classic swallowing; middle-period ones
were dilution and fabrication; late ones included \emph{a fix introducing
a deeper silence} (a retry helper whose exit-code propagation silently
killed twenty \texttt{set~-e} callers mid-script---the fix for a silent
failure became a worse one), \emph{a correct fix to the wrong bug} (an
ownership repair that was real but irrelevant to the EPERM it targeted,
exposed only because the metric refused to move), and \emph{vacuous
verification} (the 67 checks executing empty strings). The lesson we draw
is structural: silent failure cannot be enumerated and fixed; it can only
be \textbf{governed}---by meta-rules that constrain whole mechanism
families, scanners that enforce the meta-rules mechanically, and
validation that the guards themselves are alive (\S\ref{sec:defense}).

\subsection{Defense maturation: point fix $\rightarrow$ meta-rule
$\rightarrow$ scanner}\label{sec:maturation}

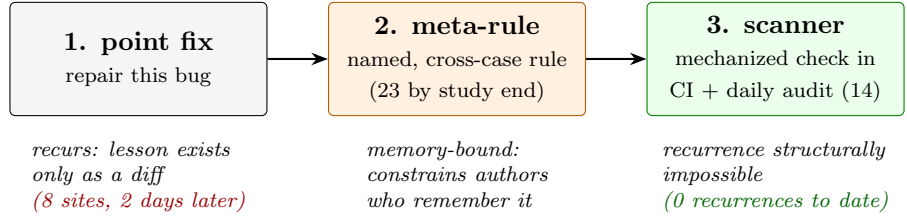
\begin{figure}[t]
\centering
\begin{tikzpicture}[node distance=3mm and 8mm]
  \node[stage, minimum width=34mm, minimum height=15mm] (fix)
    {\textbf{1. point fix}\\\scriptsize repair this bug};
  \node[amp, minimum width=34mm, minimum height=15mm, right=of fix] (rule)
    {\textbf{2. meta-rule}\\\scriptsize named, cross-case rule\\\scriptsize (23 by study end)};
  \node[good, minimum width=34mm, minimum height=15mm, right=of rule] (scan)
    {\textbf{3. scanner}\\\scriptsize mechanized check in\\\scriptsize CI + daily audit (14)};
  \draw[flow] (fix) -- (rule);
  \draw[flow] (rule) -- (scan);
  \node[lbl, below=2mm of fix]  {recurs: lesson exists\\only as a diff\\\textcolor{red!60!black}{(8 sites, 2 days later)}};
  \node[lbl, below=2mm of rule] {memory-bound:\\constrains authors\\who remember it};
  \node[lbl, below=2mm of scan] {recurrence structurally\\impossible\\\textcolor{green!40!black}{(0 recurrences to date)}};
\end{tikzpicture}
\caption{Defense maturation path. Every meta-rule that reached step~3 has
zero recorded recurrences; every recurrence in the corpus involves a
lesson that stopped at step~1.}
\label{fig:maturation}
\end{figure}

Defense effectiveness in our record correlates with how far a lesson
traveled along a three-step path (Fig.~\ref{fig:maturation}):

\begin{enumerate}
\item \textbf{Point fix} --- repair this bug. \emph{Empirically
insufficient:} an import-omission bug fixed at one site recurred at eight
sites via an automated injector \textbf{two days later}, because the
lesson existed only as a diff.
\item \textbf{Meta-rule} --- write the lesson as a named, cross-case rule
(23 such rules by study end; e.g., diagnostics-to-stderr; key-based
parsing; alert-path independence; declared-state convergence;
reserved-file unwritability). Necessary but memory-bound: rules constrain
authors who remember them.
\item \textbf{Mechanized scanner} --- encode the meta-rule as a
repository-wide check that runs in CI and the daily audit (14 such
scanners by study end: cross-OS quirks, path consistency, heredoc import
closure, monitor self-alarm compliance, cross-environment path
resolution, \dots). Only at this step does recurrence become structurally
impossible rather than culturally discouraged.
\end{enumerate}

Every meta-rule that reached step~3 has zero recorded recurrences as of
the study cutoff; every recurrence in the corpus involves a lesson that
had stopped at step~1.

\subsection{Audit is a regression engine, not a prediction
engine}\label{sec:audit}

Mid-study we audited our own defense framework against the first 15
incidents with three questions per incident: Q1---could the audit, as it
existed \emph{before} the incident, have caught it? Q2---why not?
Q3---do the guards added \emph{after} it block the same class going
forward? Results:

\begin{center}\small
\begin{tabular}{@{}lr@{}}
\toprule
\textbf{Metric} & \textbf{Value} \\
\midrule
Ex-ante prevention (Q1 fully) & \textbf{0 / 15 = 0\%} \\
Partial early warning & 2 / 15 = 13\% \\
Ex-post regression blocking (Q3 $\geq$ half) & \textbf{13 / 15 = 87\%} \\
Root cause of misses: a dimension the audit had never conceived &
12 / 15 = 80\% \\
\bottomrule
\end{tabular}
\end{center}

The 0\% is not an indictment of the audit---it is its \emph{job
description}. 80\% of misses were ``blank categories'': dimensions no
invariant had ever contemplated, which no amount of diligence within the
existing dimension set would have covered. Audits, like regression test
suites, encode the past. The honest engineering posture is to
(a)~maximize the regression rate (ours: 87\%, with sabotage validation
pushing reliability of that number), (b)~accept that prevention of
\emph{novel} mechanism classes comes from elsewhere---user-view
observation, adversarial review (``what could break that we would not
notice?''), and target-environment exposure---and (c)~measure the
\emph{conversion latency} from novel incident to mechanized guard, which
the three-step path (\S\ref{sec:maturation}) is designed to minimize. A
complementary adversarial audit (16 destruction scenarios injected
against the live repository: 10 replaying known incidents, 6 probing
suspected blind spots) scored 16/16 after the blind-spot batch drove its
own round of guard additions---a useful exercise precisely because its
first run did not score 16/16.

\section{The Defense Framework That Emerged}\label{sec:defense}

We summarize the framework not as a recommendation of its specifics but
as an existence proof of one coherent answer, with its audited scorecard
(\S\ref{sec:audit}) attached. Five pillars:

\paragraph{1. Declarative governance with mandatory depth.}
90~invariants~/ 827~checks in a YAML ontology, executed by an engine on a
daily audit cron and in CI. Each invariant declares its meta-rule
lineage, severity, and \emph{verification layers}; critical invariants
are mechanically required to verify at ${\geq}2$ layers
(declaration-level greps are not allowed to stand alone---a rule the
system enforces on itself via a meta-invariant, after single-layer greps
demonstrably failed to notice behavioral regressions).

\paragraph{2. Sabotage validation (``test the test'').}
Every new guard must be proven \emph{alive} by deliberately introducing
the violation it targets and observing it fire---then reverting. This
discipline caught, among others: guards that matched their own assertion
strings (self-referential greps), the 67 vacuous checks
(\S\ref{sec:classc}, found when a sabotage refused to fail), and tests
whose fixtures mirrored the wrong assumption (\S\ref{sec:classb}). In a
system whose primary failure class is \emph{silent}, an unvalidated guard
is indistinguishable from a vacuous one.

\paragraph{3. Declared-state convergence.}
A convergence engine diffs every declared registry (jobs, providers,
services, KB sources, runtime config) against observed runtime state on
each audit, with per-spec staged escalation: alert-only $\rightarrow$
machine-sync with dry-run default $\rightarrow$ live synchronization
after a one-week zero-drift observation window. This retired the largest
\classref{E} mechanism (operational omission) from ``human memory item''
to ``machine-closed loop''. Its own history doubles as a cautionary tale
(\S\ref{sec:discussion}): an audit that \emph{applies} synchronization is
an observer mutating the observed, and one drift bug caused a
thrice-recurring crontab duplicate until the audit path was forced to
dry-run---leading to a standing rule that \textbf{audit observes and
never mutates}.

\paragraph{4. Context hygiene and anti-fabrication layers (\classref{D}
defense).}
Mechanism-level: shell diagnostic output must go to stderr
(scanner-enforced repo-wide); alerts are tagged at every producer and
stripped from chat context at the proxy \emph{before} truncation;
reserved runtime files are unwritable by LLM tools (proxy-intercepted).
Content-level: a single shared module provides a six-level cumulative
ladder of anti-fabrication guards injected into all LLM-calling jobs
(level selection by task risk; upper levels contain literal prohibited
phrases harvested from actual fabrication incidents and tag requirements
like \texttt{[strong-evidence]}~/ \texttt{[weak-association]} on every
cross-domain claim); a source-credibility module labels every ingested
source on a five-tier provenance scale at prompt-injection time.
Post-deployment measurement showed the targeted fabrication patterns
(multi-hop causal chains, ``therefore''-style necessity claims) dropping
by 53--92\% in the affected synthesis job while tagged-claim usage went
from zero to ${\sim}9$ per day.

\paragraph{5. Monitoring that monitors itself, and alarms that outlive
their subject.}
Watchdog ERR-trap self-alarms plus heartbeat canary files age-checked
independently; alert routing that never depends on the failing subject
(gateway-down notifications travel a second transport); freshness
guarantees on every health field; and an LLM ``observer'' job that
critiques the previous day's user-facing output---the beginning of
mechanizing \S\ref{sec:discovery}'s human advantage.

\medskip
What the framework deliberately does \textbf{not} claim: prevention of
novel mechanism classes (\S\ref{sec:audit}), and net complexity
reduction---which motivates \S\ref{sec:discussion}.

\section{Discussion: Seams, Not Components}\label{sec:discussion}

Two months in, facing a week with five incidents, we asked the obvious
question: \emph{is the system failing because it has become too complex?}
The postmortem-backed answer was more specific and, we believe, more
useful: \textbf{no individual failing part was complex.} A symlink. An
\texttt{abspath} call. A one-line registry entry. A boolean default.
Every part was simple and locally correct; every incident lived in a
\emph{combination}---symlink $\times$ non-resolving path API $\times$ a
particular invocation route; sync-enabled audit $\times$ bare-name entry
$\times$ format mismatch $\times$ non-dry-run path. Components are
covered by 4{,}286 tests; \textbf{combinations grow superlinearly and
tests cover only the combinations someone imagined.} Complexity is about
parts; incidents are about seams.

This reframing has bite because it inverts the instinctive response to
incidents---\emph{add a guard}---which itself adds a part, and therefore
seams. (Our convergence engine, built to close the declared-vs-runtime
seam, opened an observer-mutates-observed seam that produced three
incidents before being caged.) The stable posture we landed on, codified
as a standing ``Sunset Law'' with two operational meta-rules, is: before
adding any mechanism, attempt to \emph{retire} an equivalent one; one
logical entity must have one physical representation (multiple
representations \emph{will} drift, and the bug lives between them); and
defenses themselves are incident surfaces---prefer \textbf{seam
reduction} (unify representations, shrink the dev--production gap, make
observers read-only) over defense accretion. In the five weeks since
adoption, the system retired more representation duplicates than it added
invariants---and we note, with appropriate caution about confounds, that
incident frequency declined while feature velocity did not.

\paragraph{On the AI-assisted operation of an AI system.}
This system is developed and operated by one human domain expert working
with an AI engineering collaborator, governing a runtime powered by
\emph{other} LLMs. Several findings above are entangled with that
arrangement in both directions. The postmortem protocol's rigidity
(mandatory causal-chain diagram before any fix; the three-question rule
before touching code) exists substantially \emph{because} an AI
collaborator will otherwise pattern-match symptoms to plausible fixes
with great speed---industrializing the trigger-level cosmetic fix
(\S\ref{sec:tac}); one documented incident chain consisted of five
cascading ``fixes'' to a problem whose correct resolution was a single
file copy. Conversely, the framework's volume (90 invariants, 14
scanners, 4{,}286 tests, 22 publication-grade postmortems in eight weeks)
is achievable for a single human \emph{only} with such a collaborator. We
offer the observation, not a verdict: AI collaboration shifts the binding
constraint of reliability engineering from implementation bandwidth to
\textbf{judgment discipline}---exactly the property the meta-rules exist
to encode. The governance framework is, in this reading, not just the
system's immune system but the collaboration's contract.

\paragraph{Generalizability.}
We claim the \emph{taxonomy classes and cross-cutting structures}
(latency $\propto$ observational distance; trigger/amplifier/concealer;
fix$\rightarrow$rule$\rightarrow$scanner maturation;
audit-as-regression) generalize to LLM agent systems with long-running
scheduled autonomy and human-facing output---the mechanisms are not
artifacts of our stack, and Classes A/C/E have direct ancestry in systems
known to fail this way at every scale. We explicitly do not claim the
\emph{frequencies} generalize (single system, single operator pair, one
OS), nor that \classref{D} frequencies transfer to systems without
synthesis-and-push pipelines. What we most hope transfers is the method:
complete causal-chain postmortems, a mechanism-oriented catalog treated
as the unit of institutional memory, and sabotage-validated conversion of
every lesson into a machine check.

\section{Threats to Validity}\label{sec:threats}

\paragraph{Construct.} ``Silent failure'' requires a silence span with
green indicators; we applied the definition at postmortem time, and
borderline cases (loud-but-cause-free alerts, \S\ref{sec:classc}) were
included when the \emph{actionable} signal was absent. Reasonable
observers could draw the line differently for 2--3 incidents.

\paragraph{Internal.} Classification was performed by the system's two
operators (human + AI) without independent annotation; we report no
$\kappa$ and acknowledge confirmation-bias risk in mechanism assignment,
partially mitigated by the classes being load-bearing (each class drives
a real scanner whose findings are objective) rather than purely
descriptive. Postmortem quality varies with log retention; two early
incidents were reconstructed retroactively under the protocol.

\paragraph{External.} One system, one host OS, one operator pair, eight
weeks, ${\sim}40$ jobs: frequencies, shares, and latencies are
descriptive statistics of a case study, not population estimates. The
70\% user-view discovery share in particular may reflect this system's
unusually attentive operator; we present it as an existence proof that
mature automated stacks can be out-detected by a user's eye, not as a
constant.

\paragraph{Survivorship/selection.} The corpus contains only failures
that \emph{stopped} being silent. Failures silent at study end are absent
by construction; the latency distribution is therefore right-censored,
and the true tail is unknown---a point that argues for, not against, the
paper's thesis.

\paragraph{AI involvement.} The AI collaborator that co-wrote the
postmortems also co-wrote this paper; the audited numbers
(\S\ref{sec:audit}, test/check counts, dates) are mechanically derivable
from the public repository, which we offer as the primary check on
narrative bias.

\section{Conclusion}\label{sec:conclusion}

Eight weeks of complete postmortems from a production LLM agent runtime
yield a five-class, mechanism-oriented taxonomy of silent failures, of
which one class---fail-plausible chained fabrication---is specific to
systems that speak. The quantified record is uncomfortable in useful
ways: a defense stack of thousands of tests and hundreds of declarative
checks prevented \emph{none} of the novel incidents ex ante (while
blocking 87\% from recurring); the best detector was a human reading the
product; and the longest failures lived not in complex code but in the
seams between simple, correct parts. The constructive program that
follows---postmortems as causal chains, lessons as meta-rules, meta-rules
as scanners, guards proven by sabotage, declared state converged by
machine, observers kept read-only, and complexity actively retired---is
less a framework than a discipline, and it is the discipline, not the
specific artifacts, that we believe transfers.

The failure your agent system should frighten you with is not the crash.
It is the confident paragraph, on schedule, in perfect grammar, about a
crisis that does not exist---pushed to a user who has every reason to
believe it, by a pipeline in which every component worked.

\section*{Artifact Availability}

All 22 incident postmortems, the canonical failure-mode catalog, the
governance ontology (90~invariants~/ 827~checks~/ 23~meta-rules), the 14
mechanized scanners, and the full test suite are public in the system
repository. The governance engine is additionally published as a
standalone, project-agnostic package on PyPI
(\texttt{openclaw-ontology-engine}, v0.1.0): any agent-runtime project
can adopt the invariant/meta-rule/scanner framework with its own YAML
configuration. A data inventory mapping every number in this paper to its
repository source accompanies the draft.

\section*{Acknowledgments}

The author thanks Claude (Anthropic) for its role as AI engineering
collaborator throughout the system's development, the incident
postmortems, and the preparation of this manuscript, as disclosed on the
title page; and the OpenClaw open-source community, whose framework the
system builds upon.


\end{document}